\documentclass[twocolumn,amsmath,preprintnumbers,superscriptaddress,amssymb, longbibliography]{revtex4-1}
\usepackage{graphicx}
\usepackage{xcolor}

\begin{document}

\title{High-Speed Control of a Fluxonium Qubit via Picosecond Pulses from Josephson Digital Circuits}

\author{M. V. Bastrakova}
\email{bastrakova@phys.unn.ru}

\affiliation{Lobachevsky State University of Nizhny Novgorod,  603022, Nizhny Novgorod, Russia}

\affiliation{Russian Quantum Center, 143025, Skolkovo, Moscow, Russia}
\affiliation{Moscow Technical University of Communications and Informatics, 111024, Moscow, Russia}
\author{D. S. Pashin}

\affiliation{Lobachevsky State University of Nizhny Novgorod, 603022, Nizhny Novgorod, Russia}
\affiliation{Moscow Technical University of Communications and Informatics, 111024, Moscow, Russia}

\author{V. A. Vozhakov}
\affiliation{Joint venture Quantum Technologies, Moscow, 115230, Russia}

\author{N. V. Klenov}

\affiliation{Moscow Technical University of Communications and Informatics, 111024, Moscow, Russia}

\author{I. I. Soloviev}

\affiliation{Russian Quantum Center, 143025, Skolkovo, Moscow, Russia}
\affiliation{Joint venture Quantum Technologies, Moscow, 115230, Russia}
\affiliation{Skobeltsyn Institute of Nuclear Physics, Lomonosov Moscow State University, 119991, Moscow, Russia}

\begin{abstract}
 We develop a control strategy for fluxonium qubits based on picosecond single-flux-quantum (SFQ) pulses generated by Josephson digital circuits. A major challenge is that SFQ protocols originally designed for transmon control become inefficient for low-frequency fluxonium. In this regime, the asymmetry of flux and charge couplings leads either to excessive leakage or excessively long gate times. We show that this limitation can be overcome by two complementary protocols. The first protocol employs high-frequency SFQ pulse packets applied through the flux channel, offering a simple, analytically transparent, and robust approach for high-fidelity single-qubit control. The second protocol combines phase-shifted flux and charge pulse sequences to enhance the resonant drive and reduce the gate duration.
 Multilevel simulations demonstrate single-qubit gate fidelities above $99.99\%$; the simpler protocol remains robust under moderate variations of qubit parameters and flux bias, while the combined protocol preserves high fidelity while substantially accelerating the gate. These results establish SFQ-based cryogenic control as a realistic route to fast, accurate, and scalable manipulation of fluxonium qubits.
\end{abstract}

\date{\today}
\maketitle

\section{Introduction}
Advances in superconducting quantum simulation  \cite{Miessen2024, Zhang2024, Rosen2024} and quantum error correction \cite{Krinner2022, Moses2024} have established superconducting qubits as a leading platform for scalable quantum processors. Since performance is limited by gate errors, further progress requires balancing qubit optimization with control complexity.

Among superconducting qubit implementations, the transmon \cite{koch2007, Schreier2008} remains the most prevalent, demonstrating high-fidelity operations. However, the transmon's weak anharmonicity limits gate speed and increases susceptibility to leakage and spectral crowding \cite{Werninghaus2021, Marques2023, Morvan2022}.
Promising alternatives to the transmon aim for long coherence, noise resilience, and high gate fidelity, with the fluxonium qubit being a leading example \cite{Manucharyan2009, Nguyen2022, Rower2024, Mazhorin2024}.
Its main advantages come from its strong anharmonicity \cite{Nguyen2019}, which effectively suppresses leakage to non-computational states, and its record-breaking coherence times \cite{Somoroff2023, Wang2025}. At the same time, the spectral properties of fluxonium qubits are determined by low characteristic frequencies. Although analog microwave control of fluxonium qubits is currently the mainstream control technique \cite{Moskalenko2022, Simakov2023}, their low transition frequencies impose stringent requirements on control-line filtering and isolation. Low-frequency technical and thermal noise must be strongly suppressed, while parasitic capacitive and inductive coupling between control lines and neighboring qubits can introduce unwanted crosstalk.

In this context, the control approach based on digital superconducting electronics \cite{Mukhanov2011, Vozhakov2022}, where single-flux-quantum (SFQ) pulse trains are generated directly within the cryogenic environment \cite{Howe2022}, represents not merely an alternative but an attractive route for unlocking the full potential of fluxonium qubits in scalable processors. While the SFQ-based approach has already demonstrated its viability for implementing single-qubit \cite{McDermott2014, McDermott2018, Leonard2019, Liu2023, Vozhakov2023, Shillito2025} and two-qubit \cite{Wang2023, Bastrakova2025, Torosov2025} operations with transmons, its adaptation to fluxonium qubits demands fundamentally new solutions \cite{bastrakova2022sub} that account for their unique low-frequency spectrum and the asymmetry between charge and flux degrees of freedom.

The goal of this work is to determine whether a fluxonium qubit can be controlled efficiently by fully digital SFQ circuitry despite its low transition frequency and the strong asymmetry between flux and charge couplings. We introduce two complementary protocols: a high-frequency flux-only SFQ scheme and a faster combined protocol based on phase-shifted flux and charge control.

Section~\ref{sec2} describes the fluxonium qubit model. Section~\ref{sec:3} examines charge- and flux-based control using regular SFQ pulse trains with a repetition rate matched to the qubit frequency. A direct adaptation of the approach used for transmons in \cite{McDermott2014, McDermott2018} proves inefficient for fluxonium, leading to either increased leakage or gate durations longer than those of conventional microwave control. The modified scheme of \cite{Lapointe-Major2025}, which introduces ``on-ramp'' and ``off-ramp'' segments, improves the fidelity but still results in relatively long gate times. These limitations motivate the alternative control strategies developed in the following sections.

Section~\ref{sec4} develops a flux-only protocol based on packets of $N$ SFQ pulses applied within each qubit period and repeated for $M$ cycles, with the pulse spacing set by a high-frequency generator. A two-level approximation is used to guide the optimization of the sequence parameters, while multilevel simulations are employed to evaluate the fidelity, leakage, and robustness of the protocol. The resulting scheme enables high-fidelity single-qubit control with gate times substantially shorter than those obtained with regular SFQ sequences.

Section~\ref{sec5} introduces a faster protocol based on combined flux--charge control. Phase-shifted SFQ pulse packets are applied through both control channels in a quadrature configuration that suppresses the leading counter-rotating contribution while enhancing the resonant drive. Because of the strong asymmetry between the flux and charge couplings, the pulse numbers $N_{\varphi}$ and $N_n$ are optimized independently. The protocol further reduces the gate duration while maintaining high fidelity and provides the single-qubit rotations required for arbitrary control in combination with virtual-$Z$ operations.

Finally, Section~\ref{sec6} discusses the practical implementation and possible extensions of the proposed SFQ control schemes, while Section~\ref{sec7} summarizes the main conclusions and outlines directions for future work.

\section{Model of the Fluxonium Qubit}\label{sec2}
The fluxonium qubit represents a promising superconducting qubit architecture that combines features of the flux qubit and the charge qubit \cite{Manucharyan2009}. Its key characteristic lies in the strong anharmonicity and characteristic flux dependence of the energy spectrum, which are achieved through a large inductance. This is typically realized using arrays of Josephson junctions \cite{Masluk2012} or superconducting nanowires \cite{Hazard2019}.

\begin{figure}[h!]
    \centering
    \includegraphics[width=1\linewidth]{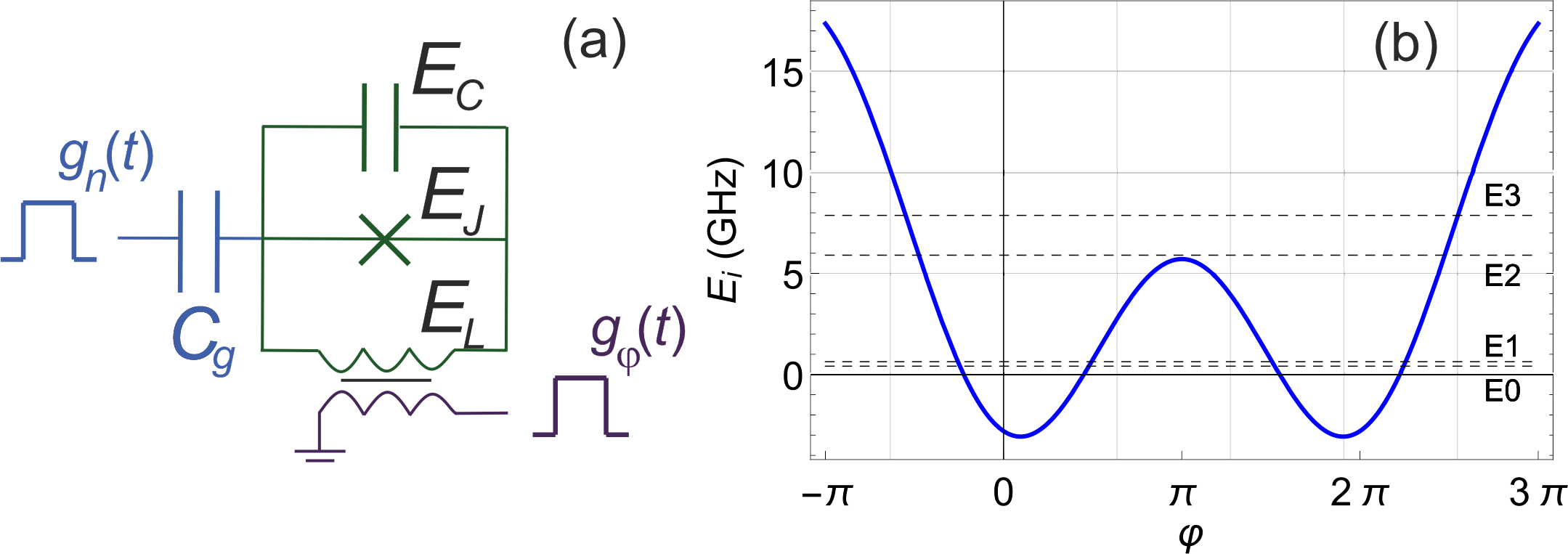}
    \caption{ (a) Schematic of the SFQ control setup and (b) fluxonium potential and low-lying energy levels at the symmetry point. The qubit parameters are as follows: $E_J/h =5.71$~GHz, $E_C/h = 1.3$~GHz, $E_L/h=0.59$~GHz. The qubit frequency is $\omega_{01}/2\pi = 222$~MHz.    
    }
    \label{fig:1}
\end{figure}

The fluxonium qubit circuit, illustrated in Fig.\ref{fig:1}(a), consists of a Josephson junction characterized by its Josephson energy $E_J = I_c\Phi_0/2\pi$, where $I_c$ is the critical current and $\Phi_0 = h/2e$ is the magnetic flux quantum. The junction also has a charging energy $E_C = e^2/2C_q$ with $C_q$ representing the junction capacitance. This junction is shunted in parallel by a large inductance $L_q$, which sets the qubit's inductive energy scale, $E_L = \Phi_{0}^{2}/4\pi^{2}L_q$.  Fluxonium typically operates with $E_J > E_C$, while the inductive shunt strongly suppresses sensitivity to static offset charge (fluctuations in the Cooper‑pair number, $n$).
The ratio $E_J/E_L$ determines the number of local wells in the qubit's potential, while the ratio $E_J/E_C$ controls the tunneling rate between these wells.  Additionally, the charging energy is typically larger than the inductive energy, $E_L \ll E_C$ \cite{Manucharyan2009}, helping to establish a strongly anharmonic spectrum and well-separated low-lying energy levels characteristic of fluxonium. The typical ratios satisfy $E_J/E_C \sim 2-10 $, and all three energies are chosen to be $E_L < E_C < E_J$. The potential profile of the fluxonium can be described as $U(\varphi)$:

\begin{equation}
U(\varphi) =  E_L\frac{(\varphi-\varphi_{\text{ext}})^2}{2} - E_J \cos\varphi, \label{eq:1}
\end{equation}
where $\varphi$ is the phase difference across the Josephson junction.

 The key operational regime of the qubit is achieved at an external flux $\Phi_{\text{ext}} = \Phi_0/2$ ($\varphi_{\text{ext}} = 2\pi \frac{\Phi_{ext}}{\Phi_{0}}= \pi$), where the potential (\ref{eq:1}) becomes a symmetric double well (see Fig.\ref{fig:1}(b)). At this point, the tunnel splitting between the low-lying basis states in the wells $\hbar \omega_{01} = E_1 - E_0$  is minimized, and its first derivative with respect to flux vanishes: $\partial\omega_{01}/\partial\Phi|_{\Phi_0/2} \approx 0$. This provides first-order protection against low-frequency magnetic flux noise and gives rise to quasi-degenerate states $|0\rangle$ and $|1\rangle$, which correspond to symmetric and antisymmetric superpositions of states localized in the individual potential minima. 
 
At the same time, the matrix element
$|n_{01}| = |\langle 0|n|1\rangle|$ of the Cooper-pair number operator remains finite, enabling charge-based control of the qubit. Thus, at the symmetry point, fluxonium combines first-order protection against flux noise with a finite charge-control matrix element.

The energy states of the qubit are described by the Hamiltonian $H_0$, expressed in terms of the canonically conjugate variables: the phase $\varphi$ and its conjugate charge $n$:
\begin{equation}
H_0 = 4E_C n^2 + U(\varphi).\label{eq:2}
\end{equation}
The typical spectrum of the fluxonium qubit, shown in Fig.\ref{fig:1}(b), is obtained by numerically solving the stationary Schrödinger equation $H_0 |j\rangle = E_j |j\rangle$, where 
$j$ denotes the level index with corresponding eigenstates $ |j\rangle $. The system parameters were chosen to be close to those in \cite{Rower2024} to enable a direct comparison between microwave and digital control methods. Figure~\ref{fig:1}(b) illustrates the strong anharmonicity characteristic of fluxonium:  $E_2-E_1 \gg E_1-E_0$, which leads to effective isolation of the computational subspace  $ \{|0\rangle,|1\rangle \}$ from higher-energy levels. 

\section{Control of the Fluxonium Qubit by Regular SFQ Pulse Trains}\label{sec:3}

We first consider regular unipolar SFQ pulse trains applied through either the capacitive (charge) or inductive (flux) control channel. The driven fluxonium Hamiltonian is
\begin{equation}
H(t) = H_0 + n A_n g_n(t)+ \varphi A_{\varphi}g_{\varphi}(t),\label{eq:3}
\end{equation}
where $H_0$ is the fluxonium Hamiltonian defined by (\ref{eq:2}), $n$ and $\varphi$ are the charge and phase operators ($[\varphi, n] = i$), respectively, $g_n(t)$ and $g_{\varphi}(t)$ are the capacitive and inductive control signals. In the simplest case, the control signal $g_{n, \varphi}(t) = f_M(t)$ is a sequence of $M$ rectangular SFQ pulses of duration $\tau$:
\begin{equation}
f_M(t) = \sum_{m=0}^{M-1}\Pi(t-mT_{01}),\label{eq:4}
\end{equation}
where $\Pi(t) = \Theta(t)\Theta(\tau - t)$, $\Theta(t)$ is the Heaviside step function, and $T_{01} = 2\pi/\omega_{01}$  is the pulse period synchronized with the qubit transition frequency. 
For an SFQ voltage pulse, the voltage-time integral is quantized,
$\int_0^\tau V(t)dt = \Phi_0$. Hence, within the rectangular-pulse approximation, the voltage amplitude is $V_0 = \Phi_0/\tau$, with the capacitive coupling to the qubit characterized by $C_g$. For the inductive control channel, we characterize the current pulse by its amplitude $I_0$, which is determined by the parameters of the SFQ driver and the control line circuit, while the inductive coupling to the qubit is characterized by the mutual inductance $L_m$. The amplitudes $A_n$ and $A_\phi$ appearing in (\ref{eq:3}) are then given by
\begin{equation}
A_{n} = \frac{C_{g}}{C_q}\, 2e V_0, \quad A_{\varphi} = \frac{L_{m}}{L_q}\, \frac{\hbar}{2e} I_0, \label{eq:5}
%A_n =\frac{C_g}{C} \int_{0}^{\tau} V(t) dt, 
\end{equation}
where $C_q$ is the qubit self‑capacitance and $L_q$ its self‑inductance. 

The single SFQ pulse induces a unitary rotation of the qubit state on the Bloch sphere. The rotation angles $\theta_{n,\varphi}$ per pulse for each control channel are respectively:
\begin{equation}
\theta_{n} = \frac{2 |n_{01}|}{\hbar}A_n \tau, \quad \theta_{\varphi} = \frac{2 |\varphi_{01}|}{\hbar}A_{\varphi} \tau,
\end{equation}
where the matrix elements $|n_{01}| = |\langle 0|n|1\rangle|$ and $|\varphi_{01}| = |\langle 0|\varphi|1\rangle|$  are determined by the computational states. We choose the relative phase of the computational
eigenstates such that $\varphi_{01}=\langle 0|\varphi|1\rangle$ is real
and positive. Using $[H_0,\varphi]=-8iE_C n$, which follows from
$[\varphi,n]=i$, this convention gives
$n_{01}=\langle 0|n|1\rangle=-i|n_{01}|$. Within the instantaneous-pulse two-level approximation, a sequence of $M$ SFQ pulses applied at successive qubit periods $T_{01}$ accumulates a total rotation angle $\theta_{\mathrm{tot}} = M\,\theta_{n,\varphi}$, thereby enabling the implementation of the desired quantum operation.

\begin{figure*}[htb]\center
    \centering
    \includegraphics[width=0.7\linewidth]{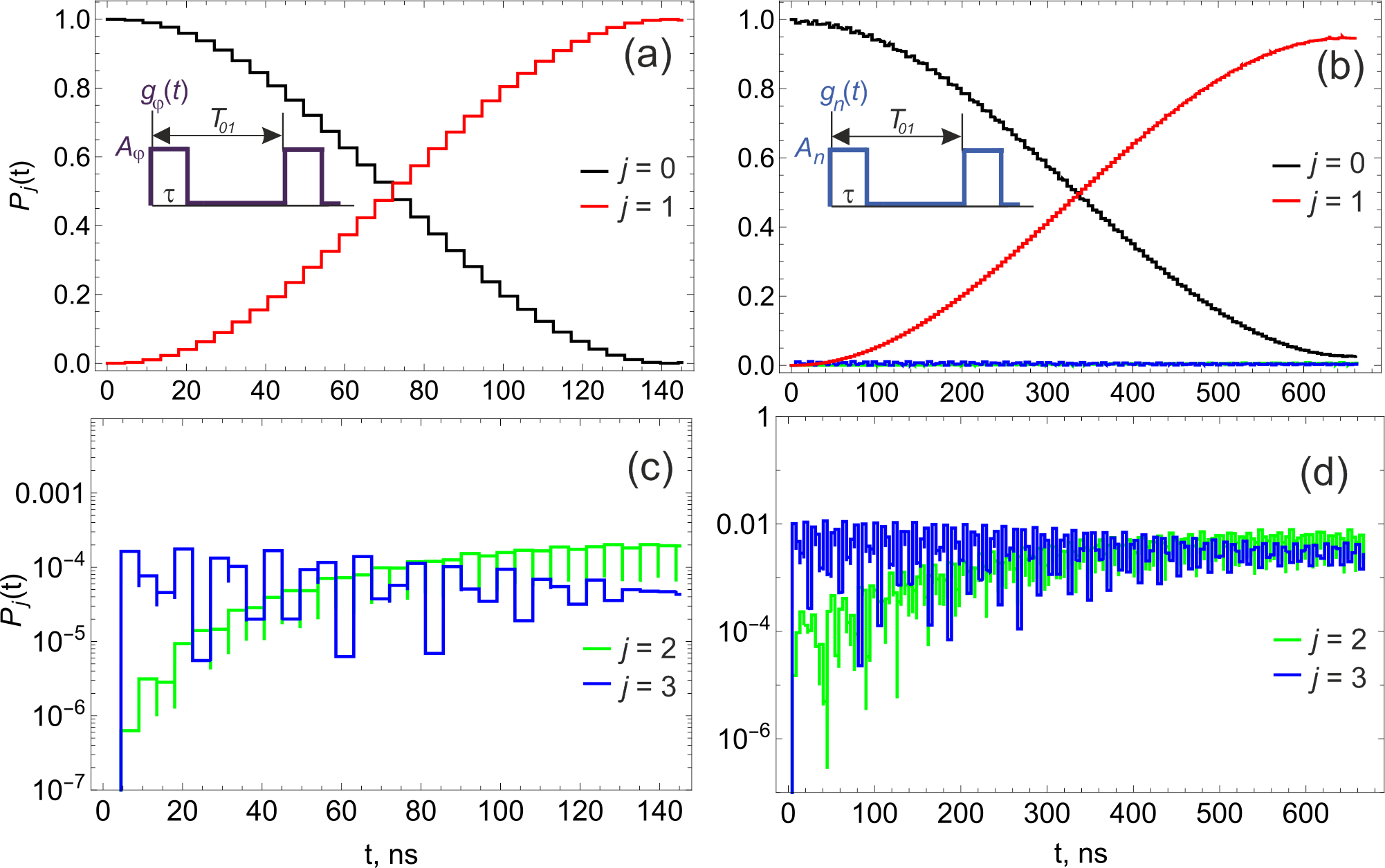}
    \caption{Dynamics of the basis state populations (a, b) and leakage (c, d) of a fluxonium qubit for the flux-control (inductive) channel (a, c) and charge (b, d) control channels using regular SFQ pulse sequences. The qubit parameters are as follows: $E_J/h = 5.71$ GHz, $E_C/h = 1.3$ GHz, $E_L/h = 0.59$ GHz, $\varphi_{\text{ext}} = \pi$.
    }
    \label{fig:2}
\end{figure*}

Figure~\ref{fig:2} shows the calculated dynamics of the fluxonium level populations for charge and flux control (\ref{eq:4}), based on the solution of the time-dependent Schrödinger equation:
\begin{equation}
        P_{j}(t) = |\langle\psi(t)|j\rangle|^2, \label{eq:7.1}
\end{equation}
where $|\psi\rangle$ is the qubit wave function. The numerical simulations were performed within a multilevel approximation taking into account 10 fluxonium levels in Eq.~(\ref{eq:3}). In this process, the number of pulses $M$ was selected to perform a qubit state flip (an $X_{\pi}$ gate) with maximum accuracy, as defined by the fidelity metric \cite{Pedersen2007}:
\begin{equation}
F=\frac{\text{Tr}( U^{\dagger}  U)+|\text{Tr}( U^{\dagger} 
 {U}_{id})|^2}{d(d+1)},\label{eq:6}
\end{equation}
The matrix $U$ entering Eq.~(\ref{eq:6}) is obtained by restricting the
full evolution operator to the computational subspace spanned by the
states $|0\rangle$ and $|1\rangle$:
\begin{equation*}
U_{ij} =
\left\langle i \left|
\mathcal{T}\exp\left[
-\frac{i}{\hbar}\int_{0}^{t_{op}} H(t')\,dt'
\right]
\right| j \right\rangle,
\qquad i,j=0,1.
\end{equation*}
Here, $t_{op}$ is the gate duration, $\mathcal{T}$ is the time-ordering operator, $U_{\rm id}$ is the ideal gate matrix, and $d=2$ is the dimension of the computational subspace. Throughout this work, the fidelity is maximized over virtual Z rotations before and after the gate.

Note that performing an $X_{\pi}$ gate  using only the flux line required $M = 32$ pulses and took $t_{op} = 145$~ns, while using only the charge line needed $M = 133$ pulses and took  $t_{op} = 600$~ns. Figures~\ref{fig:2}~(c) and (d) also reveal a significant dominance of leakage to higher noncomputational levels $|2\rangle$ and $|3\rangle$, with the leakage being an order of magnitude higher for charge-based control. Leakage from the computational subspace depends on both the control channel (charge or flux) and the number of SFQ pulses per qubit period. For example, with one pulse per period, charge control, $n$, causes more leakage than flux control, $\varphi$. For the fluxonium parameters considered here, the charge-drive matrix elements coupling the computational states to higher levels, together with the spectral content of the regular SFQ pulse train, lead to stronger excitation of noncomputational transitions than in the flux-control case. This factor has a dominant effect on the accuracy of the performed operations. The resulting gate times for both flux- and charge-based control are substantially longer, by about an order of magnitude, than those achievable with conventional microwave techniques~\cite{Nguyen2022, Rower2024}. Regular single-channel SFQ control is therefore inefficient for fluxonium: flux driving remains too slow, whereas charge driving produces substantially stronger leakage to noncomputational states. These limitations motivate restructuring the SFQ sequence within each qubit period rather than simply increasing the total number of pulses. In the following sections, we address this problem by optimizing the intra-period pulse structure, first for flux-only control and then for combined flux--charge control.

\section{Optimized flux SFQ control sequences}\label{sec4}

\begin{figure}[h!]
    \centering
    \includegraphics[width=1.0\linewidth]{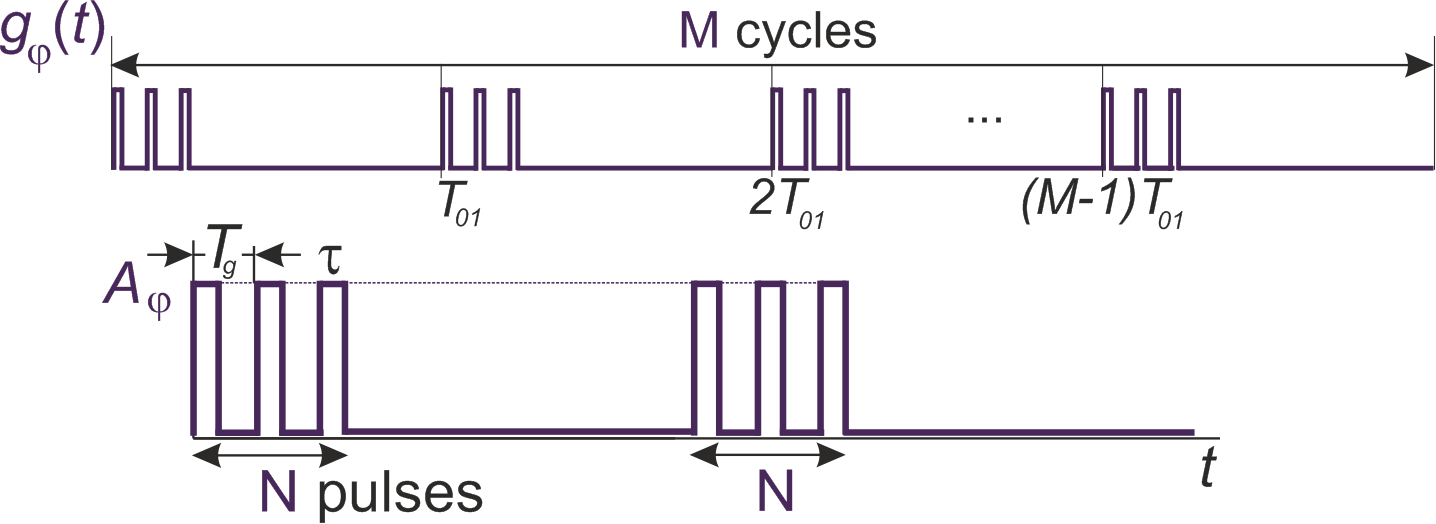}
    \caption{Control scheme based on packets of $N$ flux SFQ pulses separated by the high-frequency generator period $T_g$, with one packet applied per qubit period $T_{01}$.
    }
    \label{fig:3}
\end{figure}

In this section, we propose an approach to accelerate quantum operations using SFQ pulses applied through the flux-control channel and generated by a high-frequency source with period $T_g = 2\pi/\omega_g$, whose frequency is several tens of times higher than the qubit frequency, $\omega_g = k\omega_{01}$ ($k \gg 10$). Each pulse is modeled as a rectangular pulse of duration $\tau$ and effective flux-drive amplitude $A_\varphi$. The required SFQ pulse-generation functionality is supported by demonstrated SFQ control circuitry: in Refs.~\cite{Weng2023, Weng2024}, superconducting digital circuits based on single flux quanta have been demonstrated to generate SFQ pulse trains with adjustable parameters for qubit control. Subsequently, a packet of $N$ SFQ pulses, with the pulses separated by the generator period $T_g$, is repeated once per qubit period $T_{01}$ for $M$ cycles to perform a quantum operation. This control principle is shown schematically in Fig.~\ref{fig:3}. According to Eq.~(\ref{eq:3}), the corresponding flux-control signal can be written as $g_{\varphi}(t)=f_{M,N}(t)$, where
\begin{equation}
f_{M,N}(t) = \sum_{j=0}^{N-1}f_M(t-jT_g),  \label{eq:7}
\end{equation}
where $f_M(t)$ corresponds to expression (4).

\begin{figure}[h!]
    \centering
    \includegraphics[width=1\linewidth]{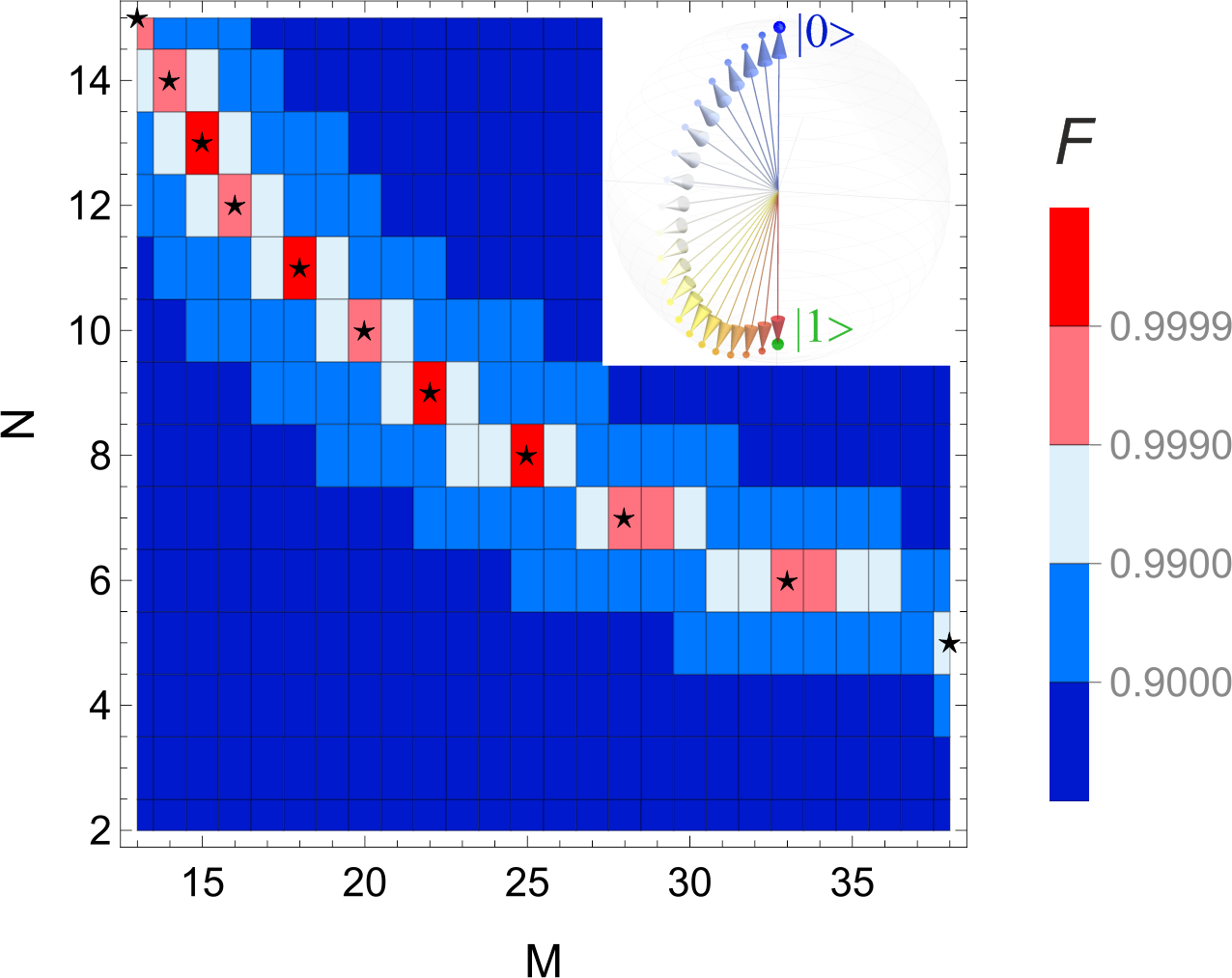}
    \caption{The color map shows the numerical simulation of the fidelity $F$ of the $X_{\pi}$ gate operation as a function of the number $N$ of SFQ pulses per qubit period $T_{01}$ and the number of repetition cycles $M$. The black markers indicate the values obtained from calculations within the two-level approximation using formula (\ref{eq:14}). The inset shows the dynamics of the qubit state vector on the Bloch sphere for $N = 9$, $M = 22$. The initial qubit state is the ground state $|0\rangle$, and the final state $|1\rangle$ demonstrates a rotation of the vector by an angle $\theta = \pi$. The qubit parameters are $E_C/h = 1.3$~GHz, $E_L/h = 0.59$~GHz, $E_J/h = 5.71$~GHz, which determine its intrinsic frequency $\omega_{01}/2\pi = 222$~MHz. The control driving is characterized by the values: $A_{\varphi}/h = 0.03$~GHz, $\tau = 15$~ps, $\omega_g = 50 \omega_{01}$.} 
      \label{fig:4}
\end{figure}

We start our analysis of SFQ pulse sequence optimization for the fluxonium (with given $A_{\varphi}, \tau, \omega_g$) by studying how the gate fidelity depends on the sequence parameters $N$ and $M$. This dependence is shown in Fig.~4, where a characteristic region (outside the blue area in Fig.~\ref{fig:4}) is observed with the fidelity reaching values $F(M,N) > 0.9$. The position of this region can be qualitatively described within the framework of the two-level approximation for the fluxonium qubit. In this approximation, projecting the full Hamiltonian (\ref{eq:3}) onto the computational subspace and retaining the flux-control term, the effective Hamiltonian $H_{\text{eff}}(t)$, taking into account the low-lying levels, is written as:
\begin{equation}
       H_{eff}(t) = \frac{\hbar\omega_{01}}{2}\sigma_z + A_{\varphi} g_{\varphi}(t)\varphi_{01}\sigma_x, \label{eq:8}
    \end{equation}
where $\sigma_x, \sigma_z$ are the standard Pauli matrices in the ordered computational basis $\{|1\rangle,|0\rangle\}$.

The evolution under the action of a single SFQ pulse, $U_{\text{SFQ}}$, and the subsequent free evolution, $U_{T_g}$, are described by the operators:
\begin{equation}
\begin{aligned}
U_{\mathrm{SFQ}} &= \exp\left[ -\frac{i}{\hbar} \left( A_{\varphi} \varphi_{01} \sigma_x + \frac{ \hbar \omega_{01}}{2} \sigma_z \right) \tau
\right],
\\
U_{T_g} &= \exp \left[ -i \frac{ \omega_{01}}{2} \sigma_z (T_g-\tau) \right].
\end{aligned}
\end{equation}
    
The full evolution over the qubit period $T_{01}$ consists of $N$ pulses synchronized with the generator frequency, occupying a total time interval $NT_g$, followed by free evolution during the residual interval $T_{01}-NT_g$. The corresponding free-evolution operator is denoted by $U_s$:
\begin{equation}
\begin{aligned}
U_{T_{01}} &= U_s \left( U_{T_g} U_{\mathrm{SFQ}} \right)^N,
\\
U_s &= \exp \left[ -i \frac{\omega_{01}}{2} \sigma_z \left(T_{01}-NT_g\right)
\right].
\end{aligned} \label{eq:12}
\end{equation}

The operator $U_{T_{01}}$ can be interpreted as a rotation of the state vector on the Bloch sphere around the axis $\vec{n}$ by an angle $\alpha$: $U_{T_{01}} = e^{-i \frac{\alpha}{2}\vec{n}\vec{\sigma}}$. The angle $\alpha$ per qubit period is calculated from the trace of $U_{T_{01}}$:
\begin{equation}
       \alpha = 2\arccos \left( \frac{|\text{Tr}(U_{T_{01}})|}{2} \right). \label{eq:13}
    \end{equation}

For the parameters considered here (flux control only, with $g_{\phi}(t)$ proportional to $\sigma_x$ in the two-level approximation), the rotation axis lies in the equatorial plane. Numerical evaluation of (\ref{eq:12}) shows that $\vec{n}$ is very close to the $X$-axis, i.e. $U_{T_{01}} \approx e^{-i\alpha \sigma_x/2}$. Therefore, each period $T_{01}$ implements a small rotation around $X$. Because the rotation axis is close to the $X$ axis, the accumulated evolution over $M$ periods approximates an $X_{\pi}$ gate. Consequently, applying $M$ such periods yields the total evolution
\begin{equation}
    U_{\mathrm{tot}} = U_{T_{01}}^{M} \approx e^{-i M\alpha \,\sigma_x/2}.
\end{equation}

To perform an $X_{\pi}$ gate, which flips $|0\rangle$ to $|1\rangle$ and vice versa, we require a $\pi$-rotation around the $X$-axis. Hence, the condition is
\begin{equation}
    M\alpha = \pi \pmod{2\pi}.
\label{eq:15}    
\end{equation}

For a given number of pulses $N$ per period, the rotation angle per period $\alpha$ is calculated via (\ref{eq:13}). The optimal integer number of cycles $M$ is then chosen as

\begin{equation}
    M = \left\lfloor \frac{\pi}{\alpha} \right\rceil,  \label{eq:14}
\end{equation}
where $\lfloor\cdot\rceil$ denotes rounding to the nearest integer. The theoretical calculation within the framework of the simplified two-level model, according to Eqs.~(\ref{eq:13}) and (\ref{eq:14}), is shown by black markers in Fig.~\ref{fig:4}. This is in good agreement with the numerical simulation and simplifies the initial estimates for selecting the required number of cycles. The evolution of the qubit state vector for the $X_{\pi}$ gate is illustrated on the Bloch sphere in Fig.~\ref{fig:4} for the values $N = 9$ and $M = 22$, where each point on the sphere represents the state rotation during each qubit period.

\begin{figure}[h!]
    \centering
    \includegraphics[width=0.8\linewidth]{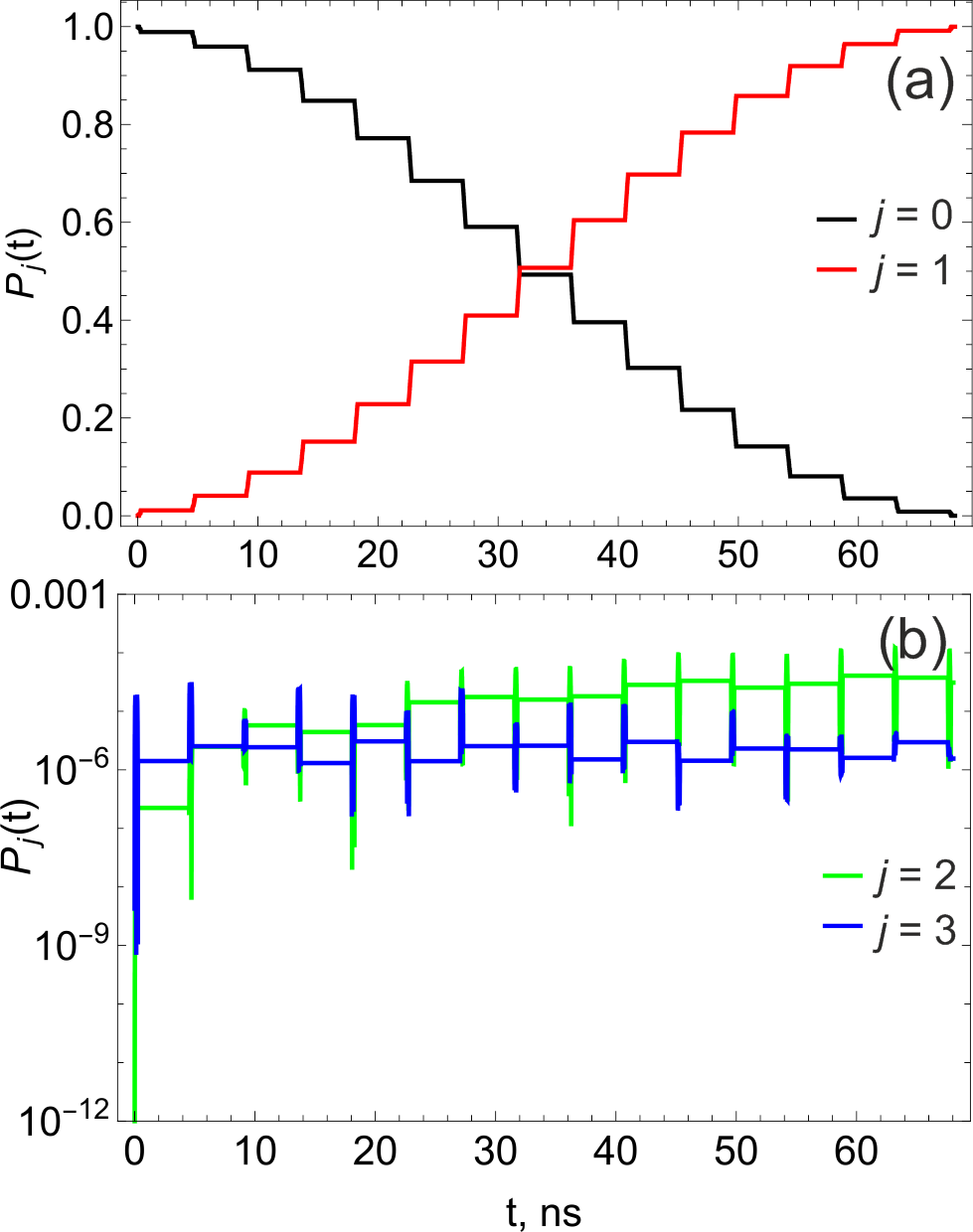}
    \caption{ Dynamics of the basis state populations (a) and leakage (b) of the fluxonium qubit for flux control based on the application of $N = 13$ pulses per qubit period, separated by $T_g = T_{01}/50$. The number of repetition cycles of the SFQ pulse sequences is $M = 15$. The remaining system and driving parameters are the same as those given in Fig.~\ref{fig:4}. 
    }
    \label{fig:5}
\end{figure}

Figure~\ref{fig:5}(a) shows the dynamics of the populations of the qubit basis states $P_{0,1}(t)$, and Fig.~\ref{fig:5}(b) shows the populations of the two higher leakage levels $P_{2,3}(t)$ in the fluxonium qubit. These results are based on numerical simulations according to Eqs.~(\ref{eq:3}) and (\ref{eq:7}) for the selected pulse parameters ($N = 13$, $M = 15$), chosen as optimal values based on the fidelity map in Fig.~\ref{fig:4} for the $X_{\pi}$ gate operation. It can be observed that leakage from the computational qubit subspace does not exceed $10^{-4}$, and the fidelity of the $X_{\pi}$ gate, calculated using formula (\ref{eq:6}), is $F = 99.995\%$. Compared with the regular flux-control sequence of Fig.~\ref{fig:2}(a), the gate duration is reduced approximately twice (and by nearly an order of magnitude compared to the charge-only case of Fig.~\ref{fig:2}(b)), while the leakage is strongly suppressed.

The strong suppression of leakage observed in Fig.~\ref{fig:5} can be understood
from the spectral structure of the SFQ pulse-packet sequence. Let $p(t)$ denote the temporal profile of a single SFQ pulse. Then, the  flux-control signal in Eq.~(\ref{eq:7}) can be written as
\begin{equation*}
g_{\varphi}(t)
=
\sum_{m=0}^{M-1}
\sum_{k=0}^{N-1}
p(t-mT_{01}-kT_g).
\end{equation*}
Its Fourier transform factorizes as
\begin{equation*}
\widetilde{g}_{\varphi}(\omega)
=
\widetilde{p}(\omega)
D_M(\omega T_{01})
D_N(\omega T_g),
\end{equation*}
where
\begin{equation*}
D_L(x)
=
\sum_{l=0}^{L-1} e^{-ilx}
=
e^{-i(L-1)x/2}
\frac{\sin(Lx/2)}{\sin(x/2)}.
\end{equation*}
The factor $D_M(\omega T_{01})$ results from the repetition of the
pulse packets with the qubit period, whereas
$D_N(\omega T_g)$ originates from the relative positions of the $N$ SFQ pulses within each packet. In contrast, for the regular SFQ sequence of Eq.~(\ref{eq:4}), containing one pulse per qubit period, the corresponding spectrum is proportional to $\widetilde{p}(\omega)D_M(\omega T_{01})$ and does not contain the additional intra-packet factor $D_N(\omega T_g)$.

Within first-order perturbation theory, the spectral weight driving a
transition $|a\rangle\rightarrow|b\rangle$ is proportional to
\begin{equation*}
\mathcal{W}_{ab}
\propto
A_{\varphi}^{2}
|\varphi_{ab}|^{2}
|\widetilde{p}(\omega_{ab})|^{2}
|D_M(\omega_{ab}T_{01})|^{2}
|D_N(\omega_{ab}T_g)|^{2},
\end{equation*}
where $\varphi_{ab}=\langle a|\varphi|b\rangle$ and
$\omega_{ab}=(E_b-E_a)/\hbar$. Thus, the effect of the packet is
determined not only by the total number of SFQ pulses, but also by
their relative phases at the frequency of each particular transition.

For the computational transition, $\omega_g=k\omega_{01}$ and therefore
\begin{equation*}
\omega_{01}T_g=\frac{2\pi}{k},
\qquad
\left|D_N(\omega_{01}T_g)\right|
=
\left|
\frac{\sin(N\pi/k)}{\sin(\pi/k)}
\right|.
\end{equation*}
For the parameters of Fig.~\ref{fig:5}, $N=13$ and $k=50$, which gives
\begin{equation*}
\left|D_{N=13}(\omega_{01}T_g)\right|
\simeq 11.61.
\end{equation*}
This value is close to the maximum possible value $N=13$, showing that the individual SFQ pulses within a packet add nearly constructively at the target transition frequency.

In contrast, the relevant leakage transitions occur at much higher frequencies because of the strong anharmonicity of fluxonium. At the symmetry point $\varphi_{\rm ext}=\pi$, the fluxonium eigenstates have definite parity and the flux operator $\varphi$ is odd. Consequently, the direct matrix element $\varphi_{02}=\langle 0|\varphi|2\rangle$ vanishes
($\varphi_{02}\simeq0$ numerically), so that direct
$|0\rangle\rightarrow|2\rangle$ excitation is suppressed. The lowest symmetry-allowed leakage channels relevant to the flux drive are therefore $|1\rangle\rightarrow|2\rangle$ and
$|0\rangle\rightarrow|3\rangle$. The former becomes relevant as population is transferred to $|1\rangle$ during the $X_{\pi}$ gate.

For the qubit parameters used in Fig.~\ref{fig:5}, the corresponding transition
frequencies are
$\omega_{12}/2\pi\simeq5.27~\mathrm{GHz}$ and
$\omega_{03}/2\pi\simeq7.46~\mathrm{GHz}$. For the same values
$N=13$ and $k=50$, the intra-packet factors are
\begin{equation*}
\left|D_{N=13}(\omega_{12}T_g)\right|\simeq0.52,
\qquad
\left|D_{N=13}(\omega_{03}T_g)\right|\simeq0.85,
\end{equation*}
which are much smaller than
$\left|D_{N=13}(\omega_{01}T_g)\right|\simeq11.61$. Thus, the pulses within each packet interfere constructively for the desired $|0\rangle\leftrightarrow|1\rangle$ transition, while they interfere predominantly destructively at the leakage frequencies. The high-frequency packet structure therefore acts as an additional spectral filter that selectively enhances the computational transition relative to unwanted transitions. The actual leakage is determined by this spectral filtering together with the corresponding matrix elements $\varphi_{ab}$ and is fully accounted for in the multilevel simulations shown in Fig.~\ref{fig:5}.

\begin{figure}[h!]
    \centering
    \includegraphics[width=1.0\linewidth]{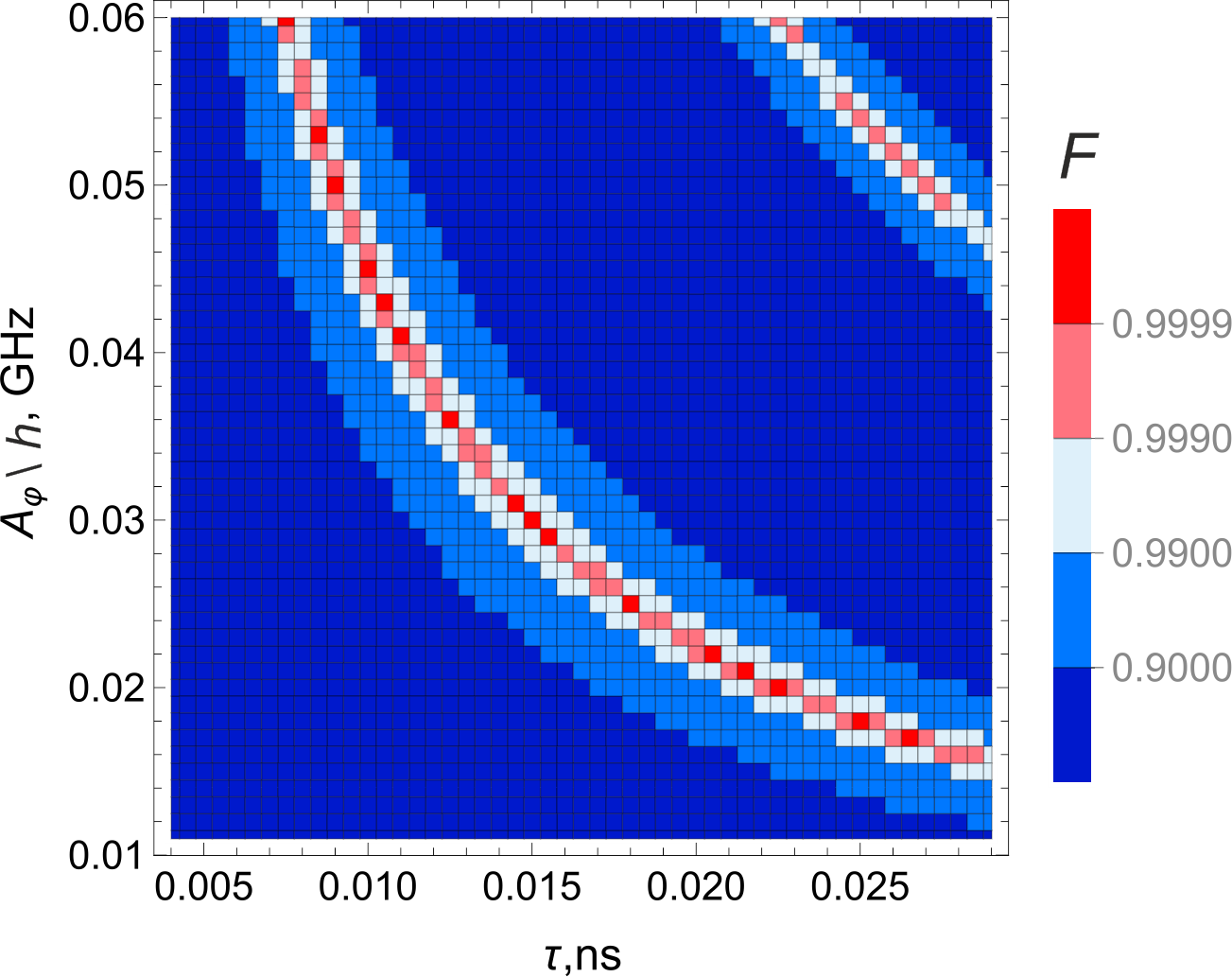}
    \caption{ The color map shows the fidelity for performing the $X_{\pi}$ gate for various amplitudes $A_\varphi$ and pulse durations $\tau$ of the drive (\ref{eq:7}). The number of SFQ pulses is $N = 13$, and the number of cycles is $M = 15$. The remaining parameters are the same as those given in Fig.~\ref{fig:4}. 
    }
    \label{fig:6}
\end{figure}

Optimizing the control drive requires choosing the optimal control signal level, defined by the SFQ pulse amplitude $A_{\varphi}$ and its characteristic duration $\tau$. Fig.~\ref{fig:6} shows that for a given pulse sequence (fixed values of $N, M$), a red ``band'' exists, where the $X_{\pi}$ gate achieves high fidelity. The high-fidelity region approximately follows $A_\varphi\tau=\mathrm{const}$, indicating that in the short-pulse regime the qubit rotation is primarily determined by the integrated effective drive strength. For $N=13, M=15$ (chosen from Fig.~\ref{fig:4}), the gate time matches Fig.~\ref{fig:5} for fixed $A_\varphi$ and $\tau$, giving  $t_{\text{op}} = MT_{01} = 68$~ns.

\begin{figure}[h!]
    \centering
    \includegraphics[width=0.85\linewidth]{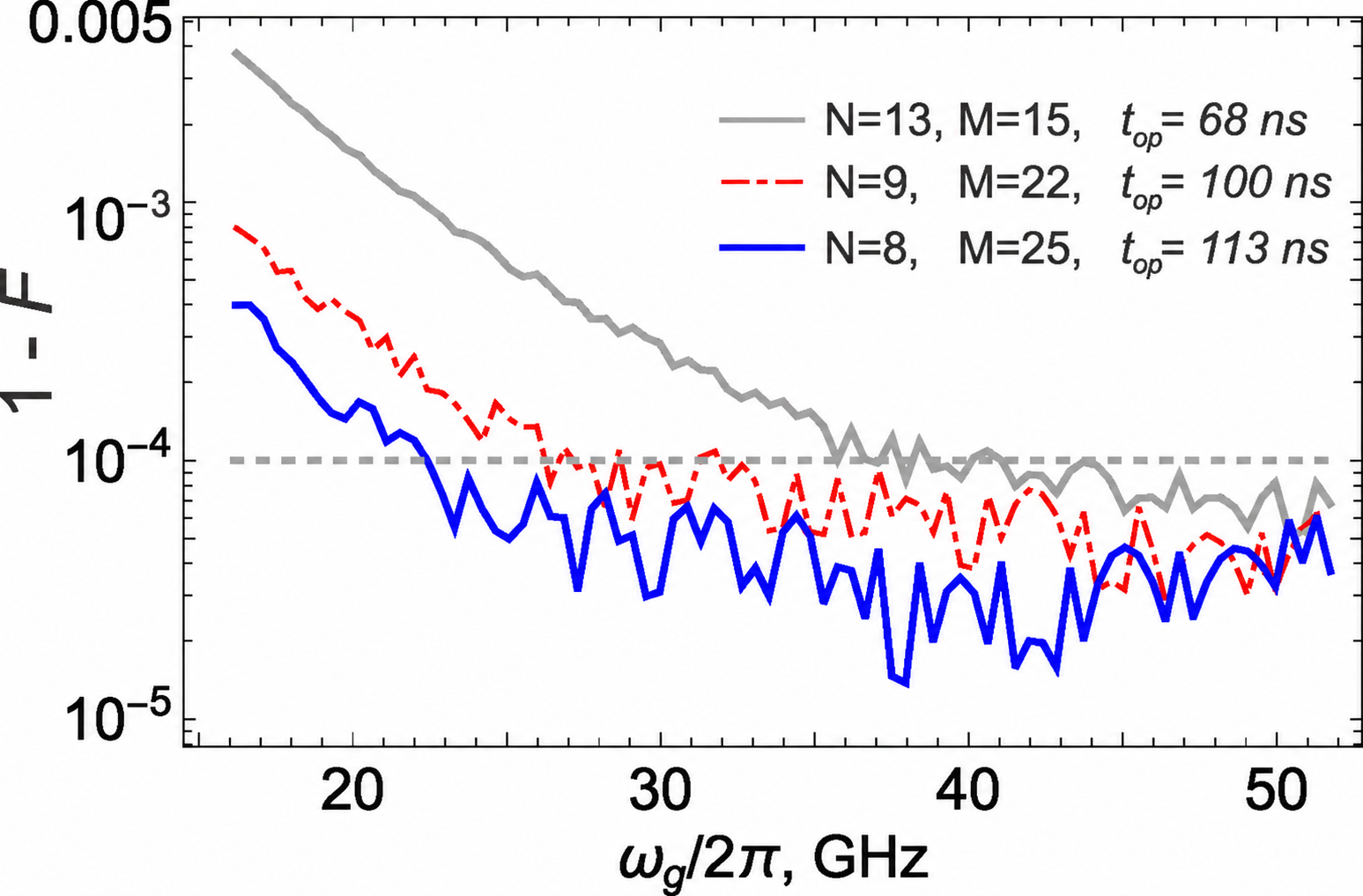}
    \caption{ Dependence of the infidelity of the $X_{\pi}$ gate on the generator frequency $\omega_g$ for different ratios of the number of pulses and repetition cycles: blue curve for $N = 8, M = 25$, red curve for $N = 9, M = 22$, and black curve for $N = 13, M = 15$. The legend shows the variation of the $X_{\pi}$ gate duration $t_{\text{op}}(\omega_g)$ for the corresponding pulse parameters. The remaining system and drive parameters are the same as those given in Fig.~\ref{fig:4}. 
    }
    \label{fig:7}
\end{figure}

After analyzing the number of control SFQ pulses (Fig.~\ref{fig:4}) and their parameters (Fig.~\ref{fig:6}), we also need to investigate the effect of the high-frequency generator frequency $\omega_g$. For clarity, Fig.~\ref{fig:7} shows the infidelity $1-F$ of the $X_{\pi}$ gate. It can be observed that as the generator frequency (the pulse repetition rate) increases, the gate fidelity improves. The legend in Fig.~\ref{fig:7} illustrates the $X_{\pi}$ gate execution time, $t_{\text{op}}$, which is primarily determined by the number of repetition cycles $M$. For each fixed $(N,M)$ sequence, increasing the generator frequency reduces the phase accumulated between successive pulses within a packet, so that their contributions become more nearly coherent at the computational transition frequency. The different $(N,M)$ combinations exhibit different sensitivities to $\omega_g$: sequences with fewer pulses per packet and a larger number of repetition cycles show a weaker dependence on the generator frequency, at the cost of a longer gate duration. Thus, the choice of $N$, $M$, and $\omega_g$ involves a trade-off between gate speed and robustness to the generator frequency. Consequently, for the  experimentally measured parameters of the fluxonium qubit and control circuitry, a preliminary analysis based on theoretical and numerical simulations is required to calibrate the selection of the required number of pulses $N$ and repetitions $M$, as well as to estimate the minimum required generator frequency $\omega_g$.

To evaluate the practical robustness of the proposed flux-control protocol based on high-frequency SFQ pulse sequences (Fig.~\ref{fig:3}), we analyze the effect of deviations in the fluxonium parameters $E_J$, $E_C$, and $E_L$, as well as in the external flux bias, on the gate performance. Variations in the Josephson energy $E_J$ modify the depths of the potential wells and the barrier height between them, while changes in the inductive energy $E_L$ affect the potential curvature. Variations in the charging energy $E_C$ alter the effective mass in the charge space and thereby affect the tunneling rates between the wells. Deviations of the external magnetic flux from the symmetry point break the symmetry of the double-well potential, shift the well positions, modify the tunneling splitting, and change the transition frequency $\omega_{01}$. They also reduce first-order protection against low-frequency magnetic noise, although the control matrix elements $n_{01}$ and $\varphi_{01}$ are only weakly affected for small deviations.

For each parameter variation, the control amplitude was recalculated using the corresponding modified circuit parameters; in particular, changes in $E_L$ were accompanied by the corresponding change in $L_q$ and hence in $A_\varphi$ according to Eq.~(\ref{eq:5}).

\begin{figure*}[htb]\center
    \centering
    \includegraphics[width=0.8\linewidth]{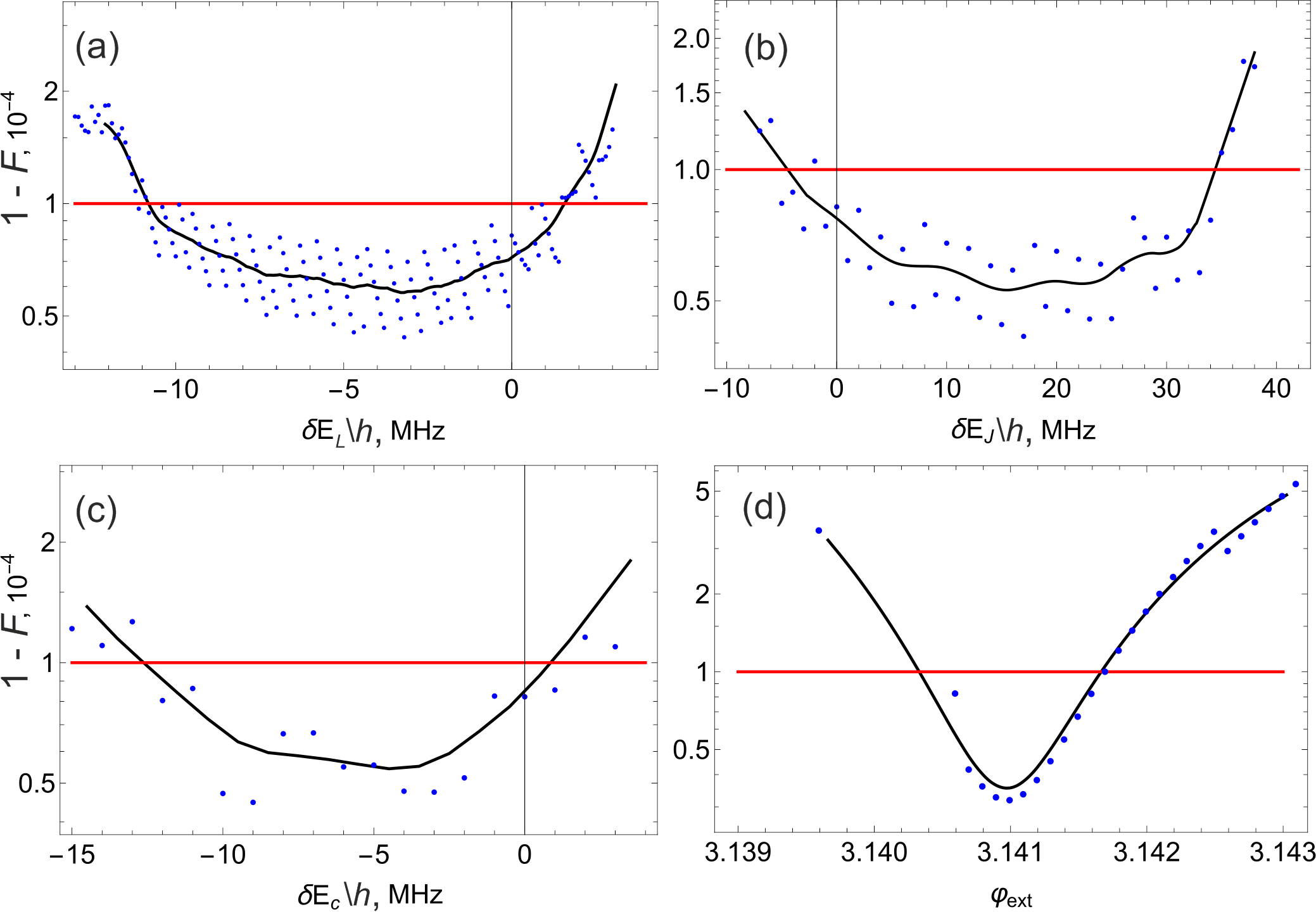}
    \caption{ Dependence of the infidelity $1-F$ of the $X_{\pi}$ gate on the deviations  of the energy parameters: (a) inductive energy $\delta E_L$, (b) Josephson energy $\delta E_J$, and (c) charging energy $\delta E_C$ from their target values ($E_C/h = 1.3$ GHz, $E_L/h = 0.59$~GHz, $E_J/h = 5.71$ GHz). Panel (d) investigates the sensitivity of the gate fidelity to the deviation of the external flux $\varphi_{\text{ext}}$ from the operating point. The system parameters and the control sequence are the same as those presented in Fig.~\ref{fig:5}. 
    }
    \label{fig:8}
\end{figure*}

For the optimized pulse sequence identified in Fig.~\ref{fig:4} ($N=13$, $M=15$) and the target fluxonium parameters $E_J/h=5.71$~GHz, $E_C/h=1.3$~GHz, and $E_L/h=0.59$~GHz, the numerical results in Fig.~\ref{fig:8} show that the $X_{\pi}$ gate fidelity satisfies the chosen criterion, $1-F<10^{-4}$, over moderate parameter variations. The protocol tolerates variations of approximately $10$~MHz in $E_C$ and $E_L$ and up to $40$~MHz in $E_J$. The same fidelity criterion is satisfied approximately within the range $3.1404 \lesssim \varphi_{\mathrm{ext}} \lesssim 3.1417$, indicating robustness to moderate deviations of the external flux bias.

Thus, the single-channel flux-packet protocol provides a simple and robust baseline for digital fluxonium control. Its further acceleration is limited by coherent drive errors, motivating the quadrature flux--charge protocol introduced in the next section.

\begin{figure}[h!]
    \centering
    \includegraphics[width=1\linewidth]{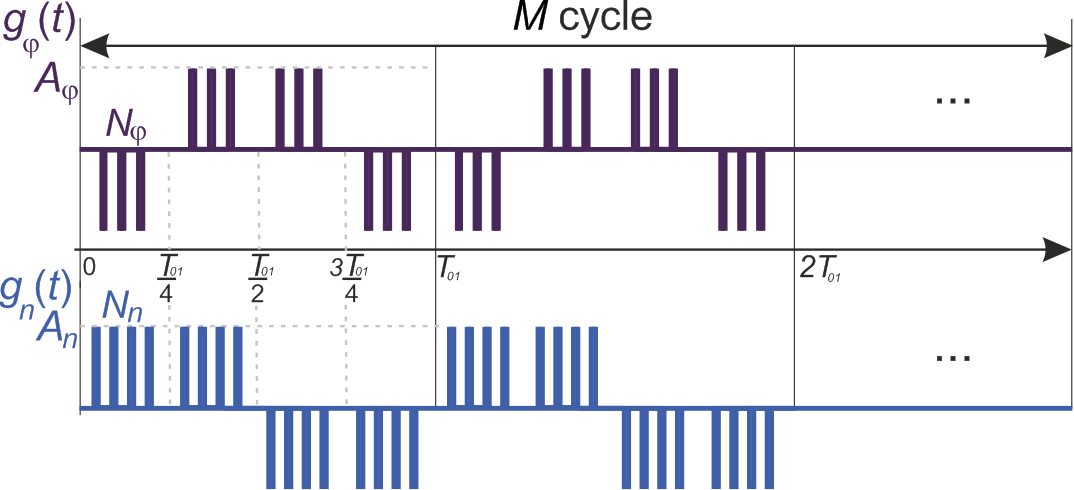}
    \caption{Schematic of counter-rotating flux-charge SFQ control of the fluxonium qubit. 
    }
    \label{fig:9}
\end{figure}

\section{Counter-rotating protocol for flux-charge SFQ control}\label{sec5} 

In this section, we adapt the method for suppressing counter-rotating errors demonstrated for microwave control in Ref.~\cite{Rower2024} to fully digital SFQ control of a fluxonium qubit. The protocol coherently drives the two canonically conjugate degrees of freedom, charge $n$ and flux $\varphi$, with a relative phase shift of $\pi/2$. In the SFQ implementation, this quadrature relation is realized by two synchronized pulse sequences applied through the charge and flux control channels, with their polarity patterns shifted by one quarter of the qubit period, $T_{01}/4$. We denote by $N_\varphi$ and $N_n$ the numbers of SFQ pulses in the flux and charge packets, respectively. For a clocked implementation, we choose $\omega_g=k\omega_{01}$ with $k/4\in\mathbb{Z}$. The packet lengths are restricted by
$$
(N_{\varphi,n}-1)T_g+\tau \leq \frac{T_{01}}{4},
$$
so that the four packets fit within one qubit period without temporal overlap.

The quarter-period synchronization is also compatible with standard clocked SFQ architectures. In particular, SFQ pattern-generator architectures can derive the control sequences from a common low-jitter clock and re-synchronize the generated bit streams before transmission to the qubit~\cite{McDermott2018}. Additional timing adjustment can be implemented at the circuit level using Josephson transmission lines; for example, Ref.~\cite{Weng2023} employs a Josephson transmission line to adjust the propagation delay within an SFQ pulse-generator circuit. Consequently, common clock fluctuations predominantly shift the two control sequences together, whereas the relevant imperfection for the present protocol is the residual differential timing error between the flux and charge channels. Previous analyses of externally clocked SFQ pulse trains have shown that timing errors need not accumulate with sequence length and can produce negligible gate errors for realistic timing jitter~\cite{McDermott2014}. A systematic quantitative study of differential timing jitter between the charge and flux channels, together with pulse distortions and other hardware-dependent imperfections, is beyond the scope of the present manuscript.

\begin{figure*}[hbt]\center
    \centering
    \includegraphics[width=1\linewidth]{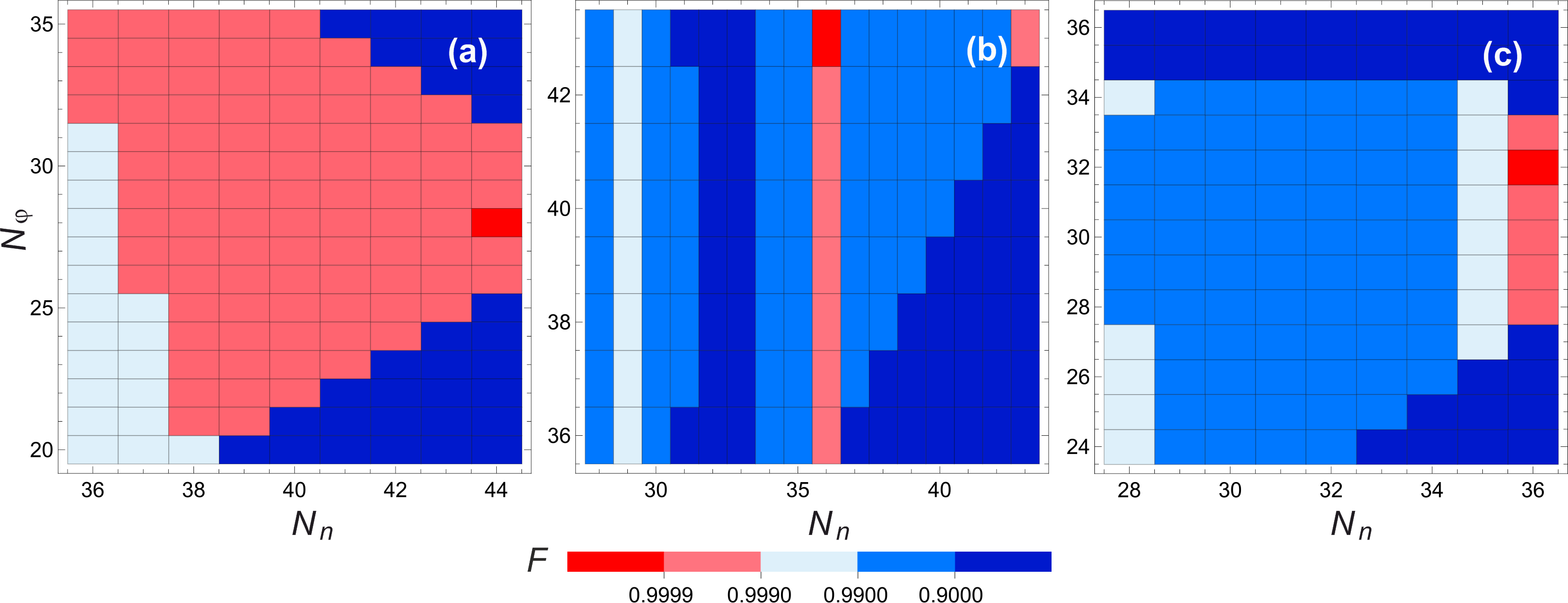}
    \caption{
    Color maps of the fidelity $F$ for the $X_{\pi}$ gate implemented using the combined flux--charge control protocol of Eq.~(\ref{eq:17}). Panels (a)--(c) correspond to Q1--Q3, respectively, and show $F$ as a function of $N_{\varphi}$ and $N_n$ at the optimized values $M=6$, $8$, and $6$, respectively. The qubit parameters and optimized effective flux-drive amplitudes are given in Table~\ref{tab}. The SFQ pulse duration is $\tau=15$~ps. The generator frequencies are $\omega_g/2\pi=40.85$~GHz, $65.41$~GHz, and $53.92$~GHz for Q1, Q2, and Q3, respectively. The same color scale is used in all panels.
    }
    \label{fig:10}
\end{figure*}

The charge control sequence, $g_n(t)$, is implemented using four SFQ pulse packets per qubit period, each containing $N_n$ pulses. Its polarity switches halfway through the qubit cycle: positive pulses are applied in the first half (at phases $0$ and $\pi/2$), and negative pulses in the second half. The flux control sequence, $g_{\varphi}(t)$, is formed by four packets containing $N_{\varphi}$ SFQ pulses each and follows the corresponding polarity pattern shifted by one quarter of the qubit period, $T_{01}/4$, relative to the charge sequence. This control scheme, illustrated in Fig.~\ref{fig:9}, is described by the expressions:
\begin{widetext}
\begin{equation}
\begin{aligned}
g_{\varphi}(t)={}&
-f_{M,N_{\varphi}}(t)
+f_{M,N_{\varphi}}\left(t-\frac{T_{01}}{4}\right)
+f_{M,N_{\varphi}}\left(t-\frac{T_{01}}{2}\right)
-f_{M,N_{\varphi}}\left(t-\frac{3T_{01}}{4}\right),
\\
g_{n}(t)={}&
 f_{M,N_n}(t)
+f_{M,N_n}\left(t-\frac{T_{01}}{4}\right)
-f_{M,N_n}\left(t-\frac{T_{01}}{2}\right)
-f_{M,N_n}\left(t-\frac{3T_{01}}{4}\right).
\end{aligned}
\label{eq:17}
\end{equation}
\end{widetext}
Here, $f_{M,N_{\varphi}}(t)$ and $f_{M,N_n}(t)$ denote the flux and charge SFQ packet sequences, respectively, with $f_{M,N_{\varphi, n}}(t)$ defined in Eq.~(\ref{eq:7}). For equal numbers of pulses in the two channels, the control amplitudes are chosen such that the effective drive strengths satisfy $A \equiv A_{\varphi}|\varphi_{01}| = A_n|n_{01}|$. In the more general case considered here, $N_{\varphi}\neq N_n$, the amplitudes are adjusted so that the integrated rotation strengths of the complete flux and charge packets are matched, $N_{\varphi}A_{\varphi}\tau|\varphi_{01}|
= N_nA_n\tau|n_{01}|$. The alternating-sign structure of Eq.~(\ref{eq:17}) requires a control circuit capable of producing both positive- and negative-polarity SFQ pulses.

With the phase convention introduced in Section~\ref{sec:3}, the flux and charge terms in Eq.~(\ref{eq:3}) drive orthogonal qubit quadratures. In particular, flux control induces a rotation about the $X$ axis, whereas charge control induces a rotation about the $Y$ axis on the Bloch sphere. For the balanced flux and charge packet strengths specified above, the quarter-period shift converts the orthogonal charge quadrature into a contribution that adds coherently to the co-rotating component of the flux drive, while the corresponding counter-rotating contributions acquire opposite phases. In the leading-order resonant approximation, these counter-rotating terms therefore cancel, as in circularly polarized microwave control~\cite{Rower2024}, while the resonant drive responsible for the qubit rotation is enhanced. For $N_{\varphi}\neq N_n$, the two pulse packets are not identical, and the cancellation is not expected to be exact; the residual contribution, together with higher harmonics and leakage, is fully included in the multilevel numerical simulations.

To illustrate the effect of the quarter-period shift independently of the optimization of $N_{\varphi}$ and $N_n$, consider first the idealized case of identical flux and charge waveforms with equal effective resonant strengths. Without the quarter-period shift, the co-rotating contributions differ in phase by $\pi/2$, so that their combined amplitude is $\sqrt{2}$ times the single-channel contribution. When the quarter-period shift is introduced, the two co-rotating contributions become phase aligned and their combined amplitude is twice the single-channel contribution. Thus, relative to the synchronous two-channel case, the quadrature configuration increases the leading-order resonant rotation rate by an additional factor of $\sqrt{2}$. For the same rotation angle, the leading-order resonant estimates of the synchronous and quadrature durations, $t_{\text{sync}}$ and $t_{\text{quad}}$, therefore satisfy
\begin{equation*}
t_{\rm sync}\simeq\sqrt{2}\,t_{\rm quad}.
\end{equation*}
This estimate does not account for higher harmonics or leakage and does not determine the speedup at a prescribed gate fidelity.

\begin{table*}[t]
\caption{
Fluxonium parameters, optimized parameters of the combined flux--charge SFQ control protocol, and calculated performance of the $X_{\pi}$, $X_{\pi/2}$, and $Y_{\pi/2}$ single-qubit gates. Here, $N_{\varphi}$, $N_n$, and $M$ are the optimized discrete control parameters for the $X_{\pi}$ gate: $N_{\varphi}$ and $N_n$ denote the numbers of SFQ pulses in the flux and charge packets, respectively, and $M$ is the number of repetition cycles. The quantity $A \equiv A_{\varphi}|\varphi_{01}|$ is the effective flux-drive amplitude. The charge-drive amplitude $A_n$ is determined from the packet-strength matching condition $N_{\varphi}A_{\varphi}|\varphi_{01}| = N_n A_n |n_{01}|$. The $X_{\pi/2}$ and $Y_{\pi/2}$ gates use the same packet parameters, with the number of repetition cycles $M$ reduced by a factor of two relative to the corresponding $X_{\pi}$ gate, and have identical gate durations and calculated fidelities within the present model.}
\label{tab}
\centering
\begin{ruledtabular}
\begin{tabular}{
c
cccc
cccc
cc
cc
}

& \multicolumn{4}{c}{Fluxonium parameters}
& \multicolumn{4}{c}{Optimized control parameters}
& \multicolumn{2}{c}{Gate duration}
& \multicolumn{2}{c}{Gate fidelity}
\\

Qubit
&
$E_J/h$
&
$E_C/h$
&
$E_L/h$
&
$\omega_{01}/2\pi$
&
$N_{\varphi}$
&
$N_n$
&
$M$
&
$A/h$
&
$t_{op_{X_{\pi}}}$
&
$t_{op_{X_{\pi/2}}}$
&
$F_{X_{\pi}}$
&
$F_{X_{\pi/2}},\,F_{Y_{\pi/2}}$
\\

&
GHz & GHz & GHz & GHz
&
&
&
&
MHz
&
ns & ns
&
&
\\
\hline

Q1~\cite{Rower2024}
&
5.71 & 1.30 & 0.59 & 0.222
&
28 & 44 & 6
&
18.753
&
27.0 & 13.5
&
0.9999003
&
0.9999896
\\

Q2~\cite{Manucharyan2009}
&
9.00 & 2.50 & 0.52 & 0.363
&
44 & 36 & 8
&
9.098
&
22.0 & 11.0
&
0.9999087
&
0.9999837
\\

Q3~\cite{Nguyen2022}
&
4.00 & 1.00 & 0.70 & 0.355
&
32 & 36 & 6
&
16.690
&
16.9 & 8.5
&
0.9999217
&
0.9999668
\\

\end{tabular}
\end{ruledtabular}
\end{table*}

Note that the quarter-period synchronization is compatible with standard clocked SFQ architectures. In particular, SFQ pattern-generator architectures can derive both control sequences from a common low-jitter clock and re-synchronize the generated bit streams before transmission to the qubit~\cite{McDermott2018}. Consequently, common-mode clock fluctuations predominantly shift the two control sequences together, whereas the relevant timing imperfection for the present protocol is the residual differential delay between the flux and charge channels. Previous analyses of externally clocked SFQ pulse trains have shown that timing errors need not accumulate with sequence length \cite{McDermott2014}. The differential timing errors between the two control channels can be eliminated by calibration of delays in dedicated Josephson transmission lines.

\begin{figure}[b!]
    \centering
    \includegraphics[width=1\linewidth]{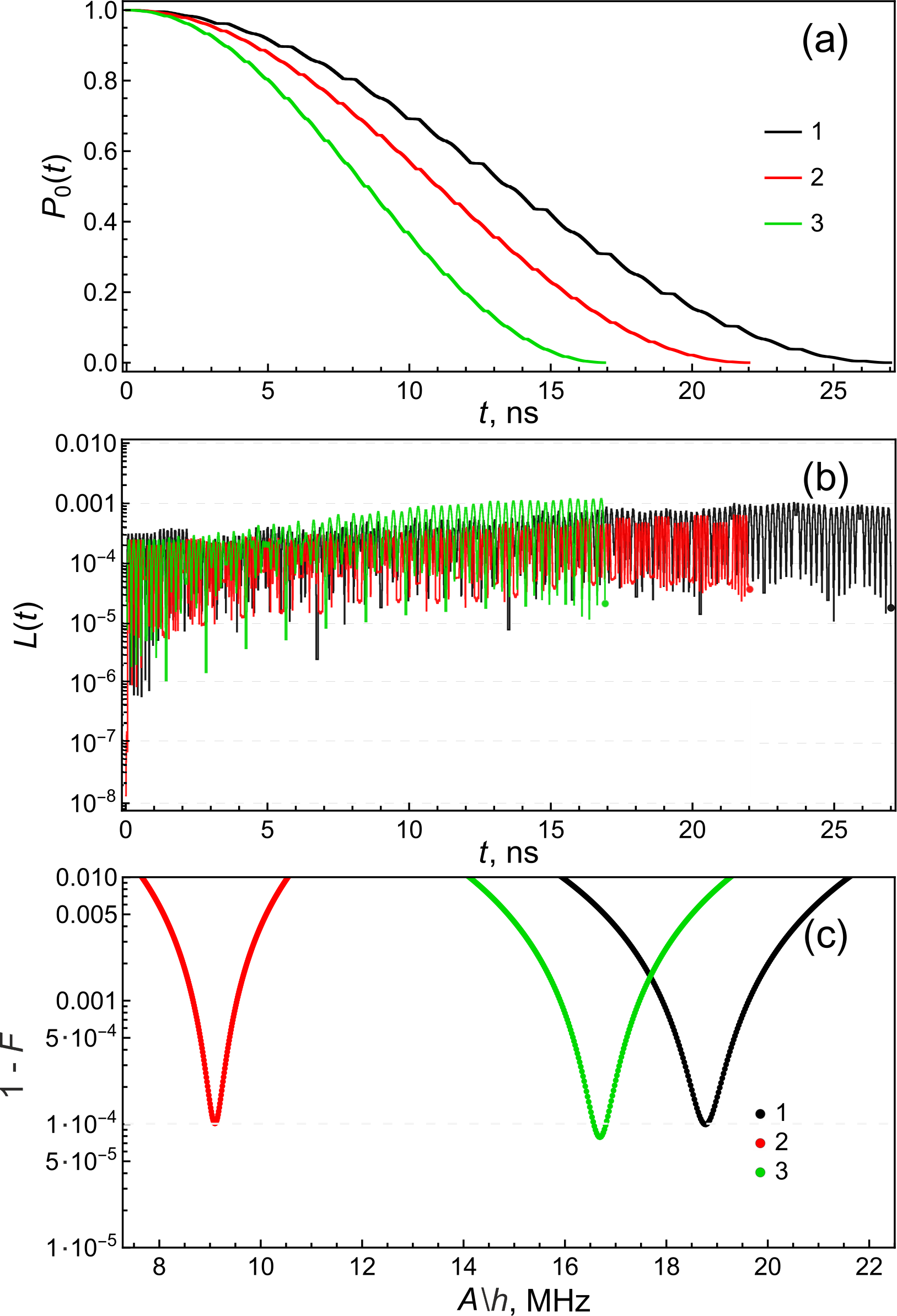}
    \caption{(a) Dynamics of the ground-state population $P_0(t)$ and (b) leakage from the computational subspace, $L(t)=1-P_0(t)-P_1(t)$, shown on a logarithmic scale for qubits Q1 (black curve), Q2 (red curve), and Q3 (green curve) when implementing the $X_{\pi}$ gate according to protocol~(\ref{eq:17}). (c) Infidelity $1-F$ as a function of the effective flux-drive amplitude $A\equiv A_{\varphi}|\varphi_{01}|$ near the optimized operating point, with $A_n$ adjusted to preserve the packet-strength matching condition $N_{\varphi}A_{\varphi}|\varphi_{01}|= N_nA_n|n_{01}|$. The pulse-packet parameters are $(N_{\varphi},N_n,M)=(28,44,6)$ for Q1, $(44,36,8)$ for Q2, and $(32,36,6)$ for Q3. The remaining parameters are the same as those given in Fig.~\ref{fig:10} and Table~\ref{tab}.    }
    \label{fig:11}
\end{figure}

The phase-shifted flux and charge pulse sequences in the flux--charge control drive of  Eq.~(\ref{eq:17}) create an effective rotating control field, enabling high-fidelity quantum operations within a fully digital SFQ control framework. To verify the performance of the combined protocol, three representative sets of fluxonium parameters were considered (Table~\ref{tab}). For each parameter set, the control sequence was optimized with respect to three discrete parameters: the number of flux pulses per packet $N_{\varphi}$, the number of charge pulses per packet $N_n$, and the number of repetition cycles $M$. The optimization was performed by a discrete numerical scan over
$N_{\varphi}$, $N_n$, and $M$ using the full multilevel evolution. For each fluxonium parameter set, the qubit parameters, pulse duration $\tau$, generator frequency $\omega_g$, and the target integrated packet strength were kept fixed. At each pair $(N_{\varphi},N_n)$, the control amplitudes $A_{\varphi}$ and $A_n$ were recalculated to preserve the packet-strength matching condition, and the highest-fidelity working point within the scanned range was selected.

Figure~\ref{fig:10} shows the fidelity of the $X_{\pi}$ gate as a function of $N_{\varphi}$ and $N_n$ at the optimized values $M=6$, $8$, and $6$ for Q1, Q2, and Q3, respectively. The optimized packet sizes satisfy $N_{\varphi}\neq N_n$ for all three parameter sets, consistent with treating the two packet sizes as independent control parameters. The optimized packet sizes, control amplitudes, gate durations, and fidelities are summarized in Table~\ref{tab}. For the $X_{\pi/2}$ operation, the number of repetition cycles is reduced by a factor of two, $M=3$, $4$, and $3$ for Q1, Q2, and Q3, respectively. 

Compared with the single-channel protocol discussed in Sec.~\ref{sec4}, the optimized combined flux--charge protocol substantially reduces the number of qubit periods required to implement the target rotation. For Q1, Q2, and Q3, the $X_{\pi}$ gate durations are approximately $27$~ns, $22$~ns, and $16.9$~ns, respectively. The corresponding $X_{\pi/2}$ and $Y_{\pi/2}$ rotations require approximately half these durations, namely $13.5$~ns, $11.0$~ns, and $8.5$~ns. At the same time, the calculated gate fidelities remain at the $99.99\%$ level or higher for all three representative fluxonium parameter sets. To limit charge-induced leakage, the required charge-control packet strength is distributed over many weakly coupled SFQ pulses, with the pulse numbers and control amplitudes optimized using the multilevel dynamics. Transient populations of higher levels may occur during the gate and can differ from the residual leakage at the gate endpoint, which is included in the fidelity.

Figure~\ref{fig:11} provides a more detailed  characterization of the optimized $X_{\pi}$ operation. Figure~\ref{fig:11}~(a) shows the ground-state population dynamics for the three fluxonium parameter sets. Figure~\ref{fig:11}~(b) presents the leakage from the computational subspace, defined as $L(t)=1-P_0(t)-P_1(t)=\sum_{j\geq 2} P_j(t)$, on a logarithmic scale. Figure~\ref{fig:11}~(c) shows the infidelity, $1-F$, as a function of the effective flux-drive amplitude $A \equiv A_{\varphi}|\varphi_{01}|$ in the vicinity of the optimized operating point. These results demonstrate that fast population transfer, strongly suppressed leakage, and a high-fidelity operating region can be simultaneously maintained for substantially different fluxonium parameter sets.

Importantly, the optimized digital control is not restricted to the $X_{\pi}$ operation. The same control architecture also provides high-fidelity $X_{\pi/2}$ and $Y_{\pi/2}$ rotations, as summarized in Table~\ref{tab}. These three operations are elements of the single-qubit Clifford group. In particular, the orthogonal $X_{\pi/2}$ and $Y_{\pi/2}$ rotations generate the complete single-qubit Clifford group, while $X_{\pi}=(X_{\pi/2})^2$.

The $Y_{\pi/2}$ operation is obtained by applying a $\pi/2$
phase shift to the corresponding $X_{\pi/2}$ control primitive. Within the present model this phase transformation introduces neither an additional gate duration nor an additional control error, which accounts for the equal calculated fidelities of the $X_{\pi/2}$ and $Y_{\pi/2}$ operations in Table~\ref{tab}. Although the Clifford group itself is not universal, combining these calibrated SFQ primitives with arbitrary virtual-$Z$ frame rotations enables the synthesis of an arbitrary single-qubit unitary. Thus, the protocol provides the physical primitives required for universal single-qubit control.

\section{Discussion}\label{sec6}

From an implementation perspective, the single-channel flux-packet protocol can be realized using standard SFQ logic elements. Figure 12 shows a conceptual implementation based on a single master SFQ clock operating at the high-frequency rate $\omega_g$. Since the packet repetition rate is synchronized with the qubit frequency, $\omega_g=k\omega_{01}$, a frequency-divider (FD) block can be used to derive the lower-frequency timing signal $\omega_{01}=\omega_g/k$ from the same master clock. The internal realization of the divider is not specified here; frequency division and multi-rate clock generation are established functions of SFQ digital circuitry. For example, SFQ frequency dividers have been used to derive several lower-frequency timing signals from a common high-frequency clock in multi-rate superconducting digital circuits \cite{frequency_division}. The sequence is controlled by two non-destructive-readout (NDRO) cells and two counters. NDRO 2 provides global control of the complete pulse sequence and is activated by the START signal. NDRO 1 controls the formation of individual pulse packets. The output pulses are simultaneously delivered to the control line and counted by Counter 1. Once $N$ pulses have been transmitted, Counter 1 resets NDRO 1, terminating the current packet, and generates a carry pulse that increments Counter 2. Counter 2 therefore counts completed packets rather than individual SFQ pulses. When the prescribed number $M$ of packets is reached, Counter 2 resets NDRO 2 and terminates the overall sequence. The values of $N$ and $M$ are loaded from the Memory Unit before the control operation.

\begin{figure}[h!]
    \centering
    \includegraphics[width=1\linewidth]{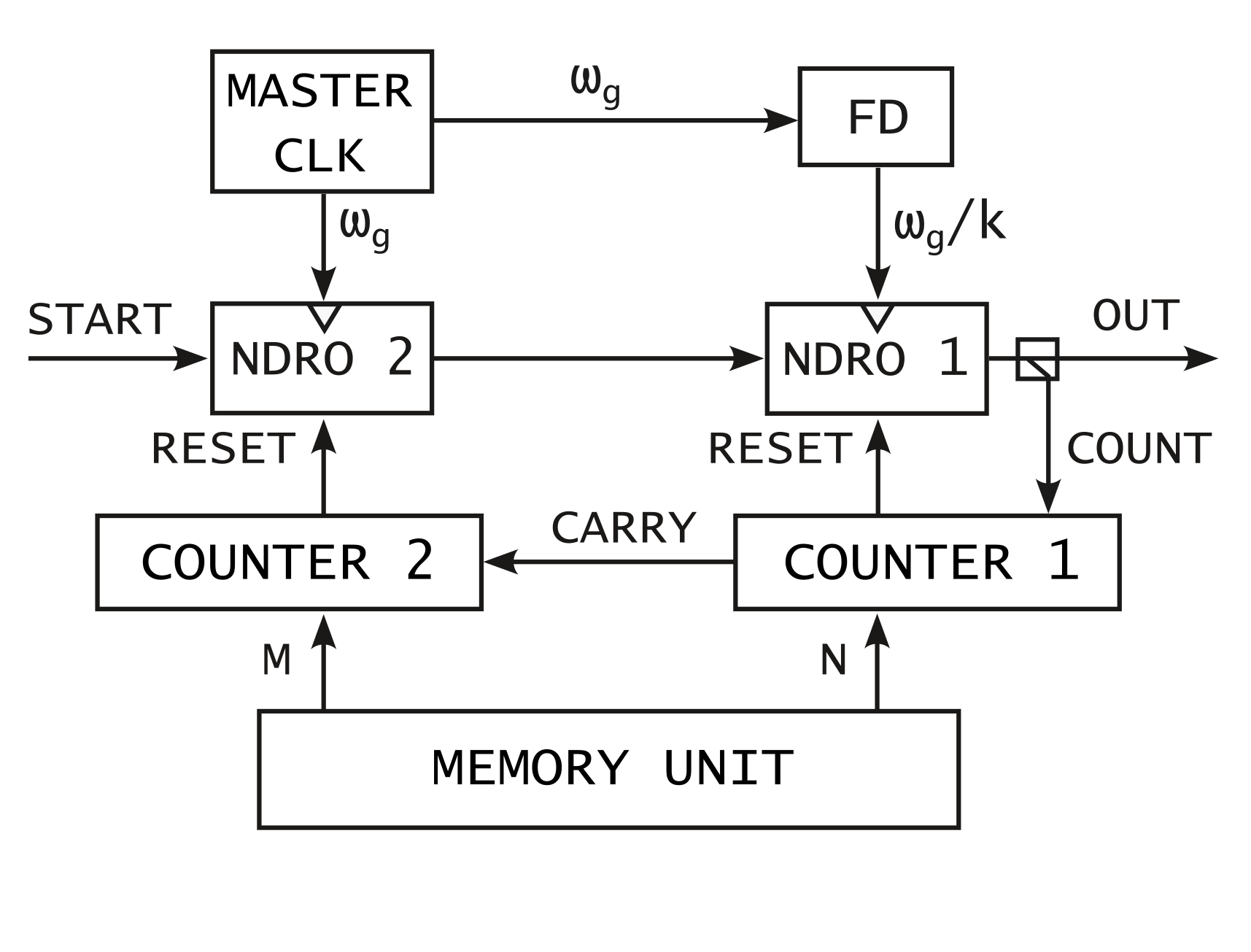}
    \caption{
     Conceptual block diagram of a possible implementation of the single-channel flux-packet SFQ control protocol. A master SFQ clock generates pulses at frequency $\omega_g$, while a frequency-divider (FD) block provides the lower-frequency timing signal $\omega_g/k=\omega_{01}$. NDRO 2 controls the overall sequence and is activated by the START signal. NDRO 1 controls the formation of individual pulse packets. Counter 1 counts the $N$ output pulses in each packet and resets NDRO 1 when the prescribed packet length is reached; its carry signal increments Counter 2. Counter 2 counts the total number $M$ of completed packets and resets NDRO 2 after the prescribed sequence length is reached. The values of $N$ and $M$ are loaded from the Memory Unit.}
    \label{fig:12}
\end{figure}

The counters may be implemented using T-flip-flop-based SFQ counter structures, while the detailed realization of the frequency-divider and pulse-distribution circuitry is beyond the scope of the present work.
The combined flux--charge protocol considered here can, in principle, be further optimized by relaxing the constraint that the integrated rotation strengths of the two control packets are matched. In the present work, this condition is imposed to reduce the dimensionality of the optimization problem and to preserve a simple balance between the two control quadratures. A more general implementation could employ independently tunable SFQ generators for the flux and charge channels, allowing the pulse repetition frequencies, packet lengths, and control amplitudes of the two quadratures to be optimized independently. Such an extended optimization may provide additional freedom to suppress residual counter-rotating errors and leakage and to further reduce the gate duration. A systematic investigation of this enlarged control parameter space is beyond the scope of the present work and is left for future studies.

\section{Conclusions}\label{sec7}
In this work, we have investigated high-fidelity, high-speed digital control of fluxonium qubits using picosecond SFQ pulses. We show that achieving such control requires protocols that substantially differ from SFQ schemes developed for transmons. Regular single-channel SFQ trains synchronized only with the qubit period either cause significant leakage or result in gate durations that are too long to be competitive. For fluxonium, this limitation arises from the combination of its low transition frequency and the strong asymmetry between the charge and flux matrix elements.

To overcome these limitations, we have developed two complementary control protocols. The first protocol uses high-frequency SFQ packets repeated once per qubit period and provides a simple and robust route to single-qubit control. For the representative fluxonium studied in Sec.~\ref{sec4}, it yields an $X_{\pi}$ gate fidelity of $F=99.995\%$ with leakage below $10^{-4}$ and a gate time of $68$~ns, while remaining stable against moderate variations of the circuit parameters and the external flux bias.

The second protocol combines independently optimized flux and charge SFQ pulse packets with a relative quarter-period shift to enhance the resonant drive. Across the representative parameter sets considered in Sec.~\ref{sec5}, the protocol realizes $X_{\pi}$ gates in approximately $17$--$27$~ns and $X_{\pi/2}$ and $Y_{\pi/2}$ gates in approximately $8.5$--$13.5$~ns, while maintaining fidelities at the $99.99\%$ level or higher.

These results demonstrate that picosecond SFQ pulses can enable fluxonium control that is both fast and precise. Next steps include extending the analysis to account for timing jitter, pulse distortions, and dissipative channels, exploring independently tunable generators for the two control quadratures without imposing the packet-strength matching condition, developing two-qubit gates within the same framework, and experimentally
benchmarking the protocols.

Overall, the integration of fluxonium qubits with digital SFQ control provides a promising framework for scalable quantum processors. By combining robust qubit operation with fast  cryogenic control, this approach offers a route toward addressing challenges in qubit manipulation and system integration.

\section{ACKNOWLEDGMENTS}
The main concept and the presented results were obtained within the framework of the Russian Science Foundation Grant No. 24-19-00187. The numerical results were obtained using the Lobachevsky supercomputer with the support of the program for the development of regional scientific and educational mathematical centers (Agreement No. 075-02-2025-1727 and Additional Agreement No. 075-02-2025-1727/1).

\bibliography{bibliography}

\end{document}